\documentclass{aa}  
\usepackage{graphicx}
\usepackage{txfonts}
\begin{document}

   \title{Energetic particle dynamics in a simplified model of a solar wind magnetic switchback}

%   \subtitle{I. Overviewing the $\kappa$-mechanism}

  \author{F. Malara \inst{1,2}
          \and
          S. Perri  \inst{1,2}
          \and
          J. Giacalone\inst{3}
          \and
          G. Zimbardo  \inst{1,2}
%          \fnmsep\thanks{Just to show the usage          of the elements in the author field}
          }

   \institute{Dipartimento di Fisica, Universit\`a della Calabria, via P. Bucci, 87036 Rende (CS), Italy\\
         \and
          Istituto Nazionale di Astrofisica, Struttura di Ricerca INAF presso Campus Universit\`a della Calabria, via P. Bucci, 87036 Rende (CS), Italy \\
         \and
          Lunar \& Planetary Laboratory, University of Arizona, Tucson, AZ 85721, USA
%             \textbf{Joe: please insert your Institution}\\
%             \email{c.ptolemy@hipparch.uheaven.space} \\
          \email{giacalon@lpl.arizona.edu}
%              \email{\textbf{Joe: please insert your email address}}
%             \thanks{The university of heaven temporarily does not                    accept e-mails}
             }
   \date{Received ; accepted }

% \abstract{}{}{}{}{} 
% 5 {} token are mandatory

\titlerunning{Energetic particles in switchbacks}
\authorrunning{Malara et al.}

  \abstract
   {Recent spacecraft observations in the inner heliosphere have revealed the presence of local Alfv\'enic reversals of the magnetic field, while the field magnitude remains almost constant. These are called magnetic switchbacks (SBs) and are very common in the plasma environment close to the Sun explored by the Parker Solar Probe satellite.}
   {A simple numerical model of a magnetic field reversal with constant magnitude is used in order to explore the influence of SBs on the propagation of energetic particles within a range of energy typical of solar energetic particles.}
   {We model the reversal as a region of space of adjustable size bounded by two rotational discontinuities. By means of test particle simulations, beams of mono-energetic particles can be injected upstream of the SB with various initial pitch- and gyro-phase angles. In each simulation, the particle energy may also be changed.}
   {Particle dynamics is highly affected by the ratio between the particle gyroradius and the size of the SB, with multiple pitch-angle scatterings occurring when the particle gyroradius is of the order of the SB size. Further, particle motion is extremely sensitive to the initial conditions, implying a transition to chaos; for some parameters of the system, a large share of particles is reflected backwards upstream as they interact with the SB. These results could have a profound impact on our understanding of solar energetic particle transport in the inner heliosphere, and therefore possible comparisons with in situ spacecraft data are discussed.}
   {}


   \maketitle
%
%-------------------------------------------------------------------

\section{Introduction}

The transport of energetic particles is a physical phenomenon found to take place in heliospheric plasmas and in other astrophysical contexts \citep{Giacalone01,Florinski03,Parizot06,Shalchi09}. Charged particles with kinetic energies ranging from tens of keV up to a few GeV, which is much higher than typical plasma thermal energies, are routinely observed in the solar wind \citep{Lee12}. Such particles can be accelerated by processes related to solar flares, coronal mass ejections, and interplanetary shocks. Due to the irregular character of the interplanetary magnetic field, the transport of energetic particles is determined by their interaction with magnetic turbulence \citep{Matthaeus03,Pucci16} and with coherent structures \citep{Tessein15} that characterise the heliospheric plasma \citep{Veltri99,Zimbardo10,Bruno13}. Particle transport is indeed affected by turbulence properties such as the fluctuation amplitude, the spectral index, and the anisotropy in the wave vector space \citep{Jokipii66,Matthaeus03,Pommois05,Hussein16,Pucci16}.
Different mechanisms determine either parallel or perpendicular transport, namely random walk of magnetic field lines, pitch-angle diffusion, and drift motion due to magnetic field inhomogeneities \citep{Moraal13,Shalchi09}. Transport properties also have implications for particle acceleration at interplanetary shock waves, because fast pitch-angle diffusion and slow spatial diffusion can speed up the acceleration process \citep{Lee82,Crooker89,Giacalone13,Amato14}.

The non-linear energy cascade process taking place in turbulence leads to the formation of coherent structures \citep{Perri12,Wu13,Greco14,Perri17,Perrone20}, which in magnetohydrodynamic (MHD) turbulence can be found as current sheets, rotational discontinuities (RDs), and tangential discontinuities (TDs). Such structures are usually observed in solar wind turbulence \citep{Tsurutani79,Borovsky10,Perri12,Greco16}, and are also related to magnetic reconnection events 
%\LEt{***Please check that I have retained your intended meaning.***}
\citep{Phan20}. In particular, RD and TD have been identified in studies based on single-spacecraft measurements \citep{Burlaga69a,Martin73,Smith73,Tsurutani79,Mariani83,Neugebauer89,Soding01} using the variance matrix method, and in studies based on multi-spacecraft observations \citep{Burlaga69b,Horbury01,Knetter03,Knetter04}. It has been suggested that the presence of RDs during Alfv\'enic periods could be related to the quasi-uniform-intensity magnetic field fluctuations \citep{Roberts12,Valentini19} that characterise such periods \citep{Belcher71}. 

The interaction of ions with RDs has been studied by \citet{Artemyev20} using a Hamiltonian formalism; these authors found that fast pitch-angle scattering is possible due to the destruction of the longitudinal adiabatic invariant. Moreover, \citet{Malara21} (hereafter, Paper I) found that ions propagating in a RD can display a chaotic behaviour where particles are temporarily trapped inside the RD, with trapping times displaying a nearly power-law distribution. RDs can actually affect energetic particles whose gyroradius is comparable to the thickness of the RD; in particular, these structures can cause fast, large pitch-angle scattering. 
%{\bf Forse la frase seguente e' superflua, il concetto e' stato gia dato sopra: Those features can affect particle transport in the presence of such magnetic structures.} 

Among the coherent structures revealed in the solar wind turbulence, magnetic switchbacks (SBs) have recently received particular attention in the literature.  A SB can be defined as a structure where the main magnetic field component ---typically the radial component $B_r$--- reverts its sign. 
 SBs have been found to exist in the interplanetary magnetic field at various heliocentric distances  \citep{McCracken66,Neugebauer13,Borovsky16,Horbury18}. Recently, in situ measurements performed by the space missions Parker Solar Probe (PSP) and Solar Orbiter (SO) showed that SBs are more frequent at shorter distances from the Sun \citep{Bale19, Fedorov21}. Information about the magnetic field line structure in SBs has been deduced by studying the propagation direction of strahl electrons \citep{Kasper19}. Such electrons move along magnetic lines, generally in the antisolar direction. The polarity reversal of $B_r$ inside a SB is typically associated with a reversal in the propagation direction of strahl electrons. This indicates that SBs can be considered as magnetic field line folds. The influence of SBs on ion propagation was studied by \citet{Bandyopadhyay21} by means of the PSP/IS$\odot$IS instrument \citep{McComas16} for energy per nucleon in the range 80--200 keV. For this energy range, \citet{Bandyopadhyay21} find that ions do not preferentially change their direction of propagation from that of the background magnetic field to that of the SBs, because of their large gyroradius.

The properties of SBs were recently examined in great detail, mainly based on PSP measurements \citep{Dudok20,Horbury20,McManus20,Mozer20,Laker21,Mozer21,Tenerani21,Pecora22}.
The changes in magnetic field direction associated with both entering and exiting a SB are quite abrupt; therefore, it can be assumed that a SB is limited by a pair of RDs where the magnetic field $\mathbf{B}$ turns in opposite ways. Other relevant properties are a correlation between the plasma velocity $\mathbf{u}$ and magnetic field $\mathbf{B}$ and a nearly constant magnetic field magnitude $B$. Therefore, SBs can be considered as (very) large-amplitude Alfv\'enic fluctuations; the propagation direction in the plasma reference frame is away from the Sun, as is the case for most of the  Alfv\'enic fluctuations observed in the solar wind. The duration of a SB, that is, the time difference in the spacecraft frame between the two crossings of the SB edges, can vary between $10^2$ s and $10^4$ s, with a distribution that follows a power law \citep{Pecora22}. Moreover, SBs are not isolated but occur in `patches' that are separated by quiet, steady wind.

The origin of SBs is still controversial. It has been proposed that SBs could originate in the solar corona as a consequence of interchange reconnection between open and closed field regions \citep{Fisk20,Bale21}. Observations of a structure reminiscent of a SB pattern propagating away from the corona in the Metis coronagraph (on board SO) data 
were recently reported \citep{Telloni22}. In addition, a local origin of SBs due to dynamical phenomena has been considered \citep[e.g.][]{Ruffolo20,Squire20,Schwadron21}. The stability and possible dissipation of SBs during their propagation has also been studied \citep{Landi06,Tenerani20,Magyar21a,Magyar21b}.

In the present paper, we study the dynamics of high-energy protons propagating across a SB, employing a simplified analytical model for the magnetic field of the SB. In particular, we investigate how the particle pitch angle is modified by the inhomogeneous magnetic field of the SB. This is accomplished by taking a test-particle approach, where single particle trajectories in phase space are determined from a numerical integration of the equations of motion. We focus on the distribution of pitch-angle variations as a function of the particle initial conditions (pitch angle, gyrophase, and energy), as well as on its dependence on parameters that characterise the SB. Chaotic features in the particle behaviour are discussed. Results are relevant in the framework of high-energy particle transport and acceleration in heliospheric plasmas.

The outline of the paper is as follows: in Sect. 2 we present a model of the magnetic field and of the dynamics of particles; in Sect. 3 we show and discuss numerical results derived from the model; finally in Sect. 4 we draw conclusions, discussing possible observations of the effects of SBs on energetic particles in the solar wind.

\section{The model}
We model a SB as a magnetic reversal included between a pair of RDs, where the magnetic field magnitude remains approximately constant. We denote the width of the RDs and the magnetic field magnitude $\ell$ and $B_0$, respectively. In particular, we assume: $B_0=1.5\times 10^{-4}$ G, a RD crossing time of $\delta t=28$ s (in the spacecraft frame of reference), and a solar wind velocity of $v_{SW}=3.4 \times 10^2$ km s$^{-1}$ \citep{Pecora22}. This gives $\ell=v_{SW}\, \delta t =9.52 \times 10^3$ km. 
However, the range of variation of these parameters can be broad \citep{Dudok20,Pecora22}. In the following, we use dimensionless quantities. In particular, the magnetic field is normalised to $B_0$ and spatial coordinates are normalised to $\ell$. Moreover, we introduce the typical proton gyrofrequency $\Omega_0 = qB_0 /(m_p c)$, where $m_p$ is the proton mass, and we normalise time to the corresponding time $\Tilde{t}=1/\Omega_0$. Particle velocity is normalised to the value $\Tilde{v}=\ell \Omega_0$. The above values of $B_0$ and $\ell$ give $\Tilde{t}=0.69$ s and $\Tilde{v}=1.37 \times 10^4$ km s$^{-1}$, respectively. We note that the value of $\Tilde{v}$ is almost equal to the speed of the protons with an energy of 1 MeV. To simplify the notation, from now on we indicate dimensionless quantities with the same symbols as the corresponding dimensional quantities, except when explicitly indicated.

%% JG Comment on the paragraph below: I think it would be reasonable to add a few sentences on the choice of a Cartesian geometry vs. a spherica one. Since the SB size (based on your assumtpions) is much smaller than the distance to the SB from the Sun (related to the field strength you assume), there is essentially nothing to be gained by going to a spherical coordinate system. It would be straightforward to relate x to the radial direction, y to the "t" direction, and z to the "n" direction in a "rtn" coordinate system. Just a thought. It is not vital to do this, but might add a bit of value to the discussion.
\subsection{Magnetic field}
We adopt a simple analytical model for the magnetic field of a SB. To represent the magnetic field $\mathbf{B}$ we use a Cartesian reference frame, indicating the corresponding unit vectors by $\mathbf{e}_x$, $\mathbf{e}_y$, and $\mathbf{e}_z$. The model is 1D, in that $\mathbf{B}$ has three non-vanishing components, but depends only on one spatial coordinate, that is, $x$: $\mathbf{B}=\mathbf{B}(x)$. 
As the SB size, as described above, is much smaller than the distance between the SB and the Sun, we can assume a local Cartesian frame without going to a spherical coordinate system. We might relate $x$ to the radial direction, as suggested by PSP observations, but more precisely $x$ represents the direction along which B most rapidly varies in space in the plasma reference frame. If the SB orientation is `oblique', the $x$ direction will not correspond to the radial.

The condition $\nabla \cdot \mathbf{B}=0$ implies that $B_x={\rm const}$. We further assume that the magnetic field magnitude is uniform: $|\mathbf{B}|=B_0={\rm const}$, where $B_0=1$ in normalised units. The above two conditions imply that $|\mathbf{B}_\perp|={\rm const}$, where $\mathbf{B}_\perp(x) = B_y(x) \mathbf{e}_y + B_z(x) \mathbf{e}_z$ is the perpendicular, variable component of $\mathbf{B}$. A magnetic field satisfying the above conditions can be written in the following form:
\begin{eqnarray}\label{B}
\mathbf{B}(x) &=& B_x \mathbf{e}_x + \mathbf{B}_\perp(x) \nonumber \\
&=& \cos \alpha \, \mathbf{e}_x + \sin \alpha \left\{ \cos \left[ \psi\left(x\right)\right] \mathbf{e}_y +
\sin \left[ \psi\left(x\right)\right] \mathbf{e}_z \right\}
,\end{eqnarray}
where $\alpha$ is the constant angle between $\mathbf{B}$ and the $x$ axis, while the quantity $\psi(x)$ represents the variable angle between $\mathbf{B}_\perp$ and the $y$ axis. We define $\alpha$ as the `obliquity angle'. We further assume that $\mathbf{B}$ is uniform both in the two regions outside the SB, where we have $\psi(x)=-\beta$, and inside the SB, where $\psi(x)=\beta$. Therefore, when crossing the two RDs that limit the SB, $\mathbf{B}_\perp$ rotates by an angle $2\beta$ and $-2\beta$, respectively. We define the constant $\beta$ as the `rotation angle'. Equation (2) is an analytical expression for $\psi(x)$ that satisfies the above assumptions:
\begin{equation}\label{psi}
\psi(x)= \beta \left[ \tanh \left(\frac{x+x_c}{\Delta x}\right) - \tanh \left(\frac{x-x_c}{\Delta x}\right) -1 \right]
.\end{equation}
According to Eq. (\ref{psi}), the two RDs are located at $x=\pm x_c$, while $\Delta x$ represents the width of the two RDs. In normalised units, $\Delta x=1$. 

\begin{figure}
   \centering
    \includegraphics[width=\hsize]{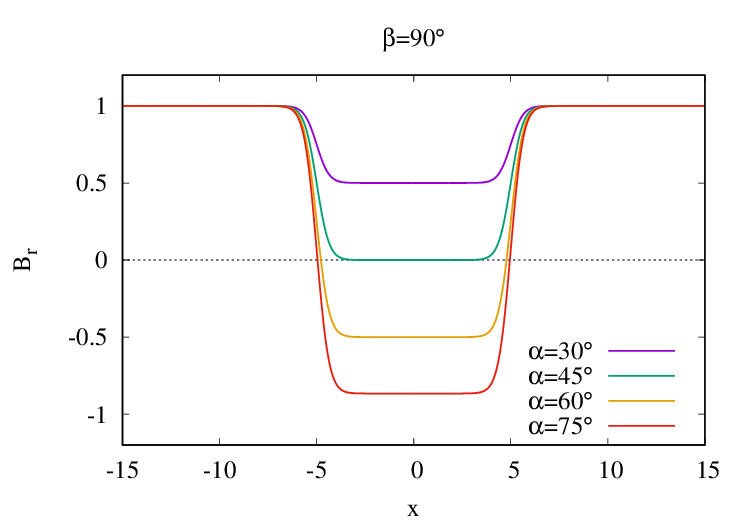}
   \caption{Radial component $B_r$ is plotted as a function of $x$, for $\beta=90^\circ$ and $\alpha=30^\circ,45^\circ, 60^\circ$, and $75^\circ$. The radial component reverts its sign in the central region for $\alpha > 45^\circ$.}
              \label{Fig:Br}%
    \end{figure}

The model for the magnetic field depends on three free parameters, namely the SB half-width $x_c$ and the two angles $\alpha$ and $\beta$. However, not all of the choices for $\alpha$ and $\beta$ correspond to a magnetic reversal. To see this, we consider the direction of the magnetic field outside the SB; conventionally, we indicate this direction as the `radial' direction. The corresponding unit vector is
\begin{equation}\label{er}
\mathbf{e}_r \equiv \lim_{x \rightarrow \pm \infty} \frac{\mathbf{B}(x)}{B(x)} =
\cos \alpha \, \mathbf{e}_x + \sin \alpha \left( \cos \beta\, \mathbf{e}_y -
\sin \beta\, \mathbf{e}_z \right)
,\end{equation}
while the `radial' magnetic field component is $B_r(x)=\mathbf{B}(x) \cdot \mathbf{e}_r$. In Fig. \ref{Fig:Br} the profile of $B_r$ is plotted as a function of $x$ for $\beta=90^\circ$ and various values of the obliquity angle $\alpha$. The figure illustrates the two RDs located at $x=\pm x_c = \pm 5$. It can be seen that, for $\beta=90^\circ$, a magnetic field reversal inside the central region is attained only for values $\alpha > 45^\circ$. 
Strictly speaking, our model can represent a SB only if the magnetic reversal is actually present, that is, if the condition $B_r(x=0) < 0$ is satisfied. 

\begin{figure}
   \centering
    \includegraphics[width=\hsize]{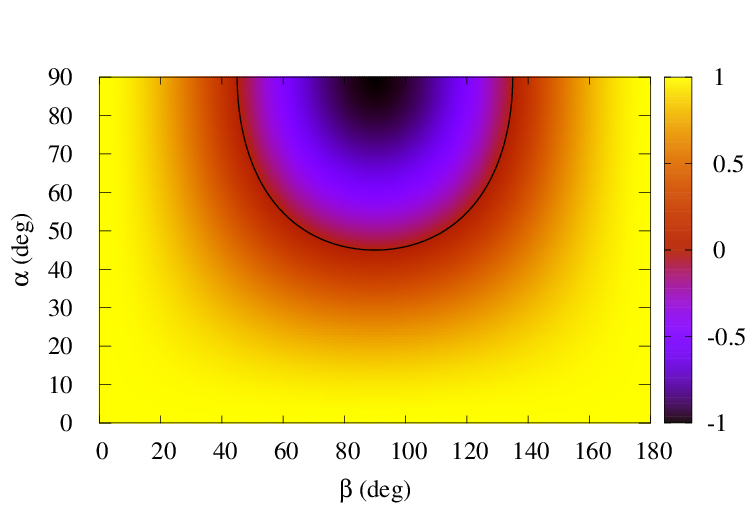}
   \caption{ $B_r(x=0)$   as a function of the angles $\beta$ and $\alpha$. The condition $B_r(x=0)<0$, corresponding to a magnetic reversal, is verified inside the blue/black coloured region limited by the black line.}
              \label{Fig:Brsign}%
    \end{figure}

In Fig. \ref{Fig:Brsign}, the value of $B_r(x=0)$ is plotted as a function of the angles $\beta$ and $\alpha$. The magnetic reversal is verified in a region of the $(\beta , \alpha)$ plane located around $\beta \sim 90^\circ$ and $\alpha \gtrsim 45^\circ$. In Fig. \ref{Fig:Brsign}, this region is
coloured blue/black and it is limited by a black line. Therefore, a SB is present when $\mathbf{B}_\perp$ performs a sufficiently large rotation across the RDs and the obliquity angle is sufficiently large. In what follows, we consider only magnetic configurations where a field reversal is present, that is, where the condition $B_r(x=0) <0$ is satisfied. The deepest reversal, where $B_r(x=0)=-1,$ is obtained for $\beta=\alpha = 90^\circ$. However, in this case we have $B_x=0$ and the two RDs become TDs. Moreover, in the whole spatial domain, magnetic field lines are straight lines contained in planes perpendicular to the $x$ direction, and therefore they do not connect the two sides of each discontinuity, as is typical of TDs. As a consequence, particles moving along field lines will not cross the discontinuities unless they are already very close to the discontinuities at the initial time. For those reasons, we do not consider the particular configuration $\alpha=\beta=90^\circ$.

\subsection{Particle dynamics}
We consider populations of protons moving in the electromagnetic field of the heliosphere. Each particle is subject to the Lorentz force $\mathbf{F}=\mathbf{F}_e + \mathbf{F}_m = q\mathbf{E} + (q/c) \mathbf{v}\times \mathbf{B}$ (in physical units), where $q$ and $\mathbf{v}$ are the proton charge and velocity; $\mathbf{E}$ and $\mathbf{B}$ are the electric and magnetic fields; and $c$ is the speed of light. It is straightforward to show that, for high-energy particles, the electric force $F_e$ can be neglected with respect to the magnetic force $F_m$. 
%Indeed, from the Faraday's law the following estimate can be derived: $\delta E \sim (\lambda/\tau) (\delta B/c) \sim (v_\phi/c) \delta B$, where $\delta E$ and $\delta B$ are the amplitudes of the electric and magnetic field fluctuations, respectively, $\lambda$, $\tau$ are the corresponding spatial and temporal variation scales, and $v_\phi$ is propagation velocity of fluctuations. Assuming that $v_\phi \sim v_A$, with $v_A$ the Alfv\'en velocity, we have:
%\begin{equation}\label{forceratio}
%\frac{F_e}{F_m} \sim \frac{\delta E}{v \delta B/c} \sim \frac{v_A}{v}
%\end{equation}
%We consider protons with a kinetic energy $\mathcal{E} \ge 10^2$ keV. This corresponds to a proton velocity $v \gtrsim 4.4 \times 10^8$ cm s$^{-1}$. Assuming for the Alfvén velocity the value $v_A \sim (3-5) \times 10^6$ %cm s$^{-1}$, typical of the solar wind, from Eq. (\ref{forceratio}) we obtain
%$F_e/F_m \le 10^{-2}$. 
%Indeed, using the Faraday's law the following relation can be derived {\bf in the reference frame at rest with the plasma}:
%\begin{equation}\label{forceratio}
%\frac{F_e}{F_m} \sim \frac{\delta E}{v \delta B/c} \sim \frac{v_A}{v}
%\end{equation}
%where we assumed that the typical value for the propagation velocity of fluctuations is given by the Alfv\'en velocity $v_A$. 
Assuming the value $v_A \sim 30-50$ km s$^{-1}$ and considering protons with velocity $v \gtrsim 4.4 \times 10^3$ km s$^{-1}$ (corresponding to energy $\mathcal{E} \ge 10^2$ keV), we obtain $F_e/F_m \le 10^{-2}$.
%Therefore, when describing the dynamics of such particles the electric force $F_e$ can be neglected with respect to $F_m$. 
%JG Comment: Would tt be difficult to just use the ideal MHD limit of E? That is: E = -V_sw x B/c? I take it that your field model has the x direction to be that of the solar wind flow - is this right? Including this field essentially just means that your calcuation is performed in the inertial frame.
For the same reason, we can neglect the time dependence in the magnetic field $\mathbf{B}$. The above conditions are met to an increasing extent with increasing particle energy. Within this approximation, the motion equations can be written in the following dimensionless form:
\begin{equation}\label{motioneq1}
\frac{d\mathbf{r}}{dt} = \mathbf{v}
,\end{equation}
\begin{equation}\label{motioneq2}
\frac{d\left[ \gamma(v)\mathbf{v}\right]}{dt} = \mathbf{v}\times \mathbf{B}
,\end{equation}
where $\gamma(v) = \left[ 1 - (v^2/c^2)\right]^{-1/2}$ is the Lorentz factor.
Equation (\ref{motioneq2}) implies that $v=|\mathbf{v}|={\rm const}$, that is, the particle energy is conserved. 
%To prove this we perform a dot product of both sides of Eq. (\ref{motioneq2}) by $\mathbf{v}$ and expand the time derivative, obtaining:
%\begin{equation}\label{vconst1}
%\frac{d \gamma(v)}{dt} v^2 + \gamma(v) \frac{d}{dt} \left(\frac{v^2}{2}\right)=0
%\end{equation}
%The time derivative of $\gamma(v)$ is $d\gamma(v)/dt=(v/c^2)\gamma^3(v)(dv/dt)$. Inserting this expression in Eq. (\ref{vconst1}) we obtain:
%\begin{equation}\label{vconst2}
%\gamma(v) v \left[\frac{v^2}{c^2}\gamma^2(v) + 1 \right] \frac{dv}{dt}=0
%\end{equation}
%For $v\ne 0$, Eq. (\ref{vconst2}) is satisfied only if $dv/dt=0$. 
This condition also implies that $\gamma(v)={\rm const}$. Therefore, the motion equation (\ref{motioneq2}) can be rewritten in a simpler form:
\begin{equation}\label{motioneq2b}
\frac{d\mathbf{v}}{dt} = \frac{\mathbf{v}\times \mathbf{B}}{\gamma(v)}
.\end{equation}
Motion equations (\ref{motioneq1}) and (\ref{motioneq2b}) are numerically integrated by employing the Boris method. It has been shown that this method is symplectic and conserves the particle energy up to the round-off error \citep{Webb14}. We note that we choose to adopt a Cartesian geometry in the plasma frame because we are integrating the particle equations in the vicinity of the SB, which means that we can neglect the curvature of the background magnetic field as well as adiabatic cooling.

% JG Comment: I think it is worth poining out that your calculations are run for such a short time that it is reasonable to consider the geometry to be Cartesian, and not spherical (like real switchbacks). In the spherical geometry, there would be adiabatic cooling of the particles and the energy is not constant. Also, if one includes the motional electric field, that I mentioned above, then it is the energy in the plasma frame that is constant. It would not be constant in the inertial frame.

\begin{figure}
   \centering
    \includegraphics[width=\hsize]{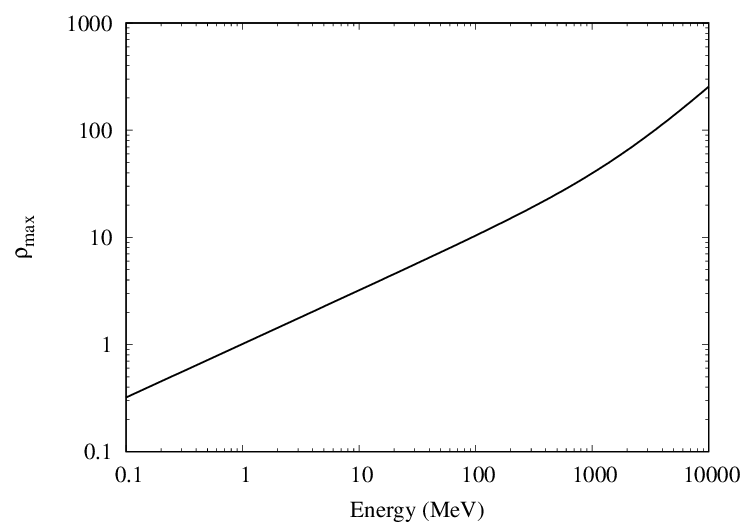}
   \caption{ Larmor radius $\rho_{\rm max}$, corresponding to $\mu=0$, is plotted as a function of the particle energy $\mathcal{E}$.}
              \label{Fig:rho_vs_E}%
    \end{figure}

We study the evolution of a population of protons that propagate across the SB. The population is characterised by a given value of the energy $\mathcal{E}$, and the corresponding proton  velocity is
\begin{equation}\label{Etov}
v = \frac{c}{\Tilde{v}}\left\{ 1 - \left[1 + \frac{\mathcal{E}}{m_p c^2}\right]^{-2} \right\}^{1/2}
,\end{equation}
where $v$ and $\mathcal{E}$ are expressed in normalised and in physical units, respectively. The velocity components parallel and perpendicular to $\mathbf{B}$ are given by $v_{||}=\mathbf{v}\cdot \mathbf{B}/B$ and $v_\perp=|\mathbf{v}-v_{||}\mathbf{B}/B|$, respectively. The ratio $\mu(t)=v_{||}(t)/v$ represents the cosine of the particle pitch angle. The particle Larmor radius is $\rho=\gamma(v) v_\perp/B$ (in normalised units), and is related to $\mu$ by $\rho=\gamma(v) v \sqrt{1-\mu^2}/B$. We define the constant quantity $\rho_{\rm max}=\gamma(v) v/B$, corresponding to the Larmor radius calculated for $\mu=0$. The Larmor radius $\rho_{\rm max}$ is plotted in Fig. \ref{Fig:rho_vs_E} as a function of the particle energy $\mathcal{E}$, for $B=1$. The dependence of $\rho_{\rm max}$ on $\mathcal{E}$ departs from being linear for energies $\mathcal{E}\gtrsim 100$ MeV.
    
Each particle is initially located outside the SB on the side of negative $x$ and moves towards the SB. The initial position for all particles is $(x_0,y_0,z_0)$, with $x_0=-(6 x_c+2 \rho_{\rm max})$ and $y_0=z_0=0$. The value chosen for $|x_0|$ is large enough to guarantee that the particle is initially well outside the SB. As the magnetic field depends only on $x$, the values of $y_0$ and $z_0$ are not relevant.

To define the initial velocity, we consider another reference frame $\left\{ x', y',z' \right\}$, where the $z'$ axis is parallel to $\mathbf{B}(x_0)$. With respect to this reference frame, the initial velocity is defined as:
\begin{equation}\label{v0}
\mathbf{v}_0 = v \left[ \sqrt{1-\mu_0^2} \left( \cos \phi_0 \mathbf{e}_{x'} + \sin \phi_0 \mathbf{e}_{y'}\right) + \mu_0 \mathbf{e}_{z'} \right]
,\end{equation}
where $\mu_0=\mu(t=0)$ is the initial pitch-angle cosine and $\phi_0$ is the initial gyrophase. The values of $\mu_0$ and $\phi_0$ are chosen in the intervals $0 < \mu_0 \le 1$ and $0\le \phi_0 < 2\pi$. Only strictly positive values of $\mu_0$ are considered, corresponding to particles initially moving toward the SB. In Eq. (\ref{v0}), the unit vector along the $z'$ axis is $\mathbf{e}_{z'} = \mathbf{B}(x_0)/B(x_0)$; the unit vector along the $y'$ axis is chosen to be perpendicular both to the $x$ axis and to $\mathbf{e}_{z'}$: $\mathbf{e}_{y'}=\left( \mathbf{e}_x \times \mathbf{e}_{z'} \right) / |\mathbf{e}_x \times \mathbf{e}_{z'}|$; finally, $\mathbf{e}_{x'}=\left( \mathbf{e}_{y'} \times \mathbf{e}_{z'} \right) / |\mathbf{e}_{y'} \times \mathbf{e}_{z'}|$. 
Using the above definitions, the initial velocity components with respect to the $\left\{ x,y,z \right\}$ reference frame are given by
 \begin{equation}\label{v0_comp}
 \begin{cases}
 v_{0x} = -\displaystyle{\frac{B_y(x_0)}{B(x_0)}} v \cos \phi_0\, \sqrt{1-\mu_0^2} + \frac{B_x(x_0)}{B(x_0)} v \mu_0 \\
 v_{0y} = \displaystyle{\frac{B_x(x_0) B_y(x_0)}{B(x_0) B_{yz}(x_0)}} v \cos \phi_0\, \sqrt{1-\mu_0^2} \\
\;\;\;\;\;\;\; -\displaystyle{ \frac{B_z(x_0)}{B_{yz}(x_0)}} v \sin \phi_0\, \sqrt{1-\mu_0^2} +
 \frac{B_y(x_0)}{B(x_0)} v \mu_0 \\
 v_{0z} = \displaystyle{\frac{B_x(x_0) B_z(x_0)}{B(x_0) B_{yz}(x_0)}} v \cos \phi_0\, \sqrt{1-\mu_0^2} \\ 
\;\;\;\;\;\;\; + \displaystyle{\frac{B_y(x_0)}{B_{yz}(x_0)}} v \sin \phi_0\, \sqrt{1-\mu_0^2} +
 \frac{B_z(x_0)}{B(x_0)} v \mu_0.  
 \end{cases}
 \end{equation}

Time integration is carried out until the given particle has completely left the SB, either on the positive or negative $x$ side (the latter situation corresponding to a particle that is reflected back by the SB). This condition is well met numerically when $|x(t)| > |x_0| + 2\rho_{\rm max}$.

\section{Results}
In the problem under study, the particle energy is conserved in time. Therefore, the effect of the inhomogeneous magnetic field on particles is an energy transfer from parallel to perpendicular motion, or vice versa. This is equivalent to a change in the pitch angle cosine $\mu$. Considering a population of protons, this can correspond to a scattering in pitch angle or to a focusing process, as we show below. The details of this process depend
%\LEt{***Please check that I have retained your intended meaning.***}
 on the parameters defining the magnetic field structure and on the particle energy. In this section, we discuss numerical results and their dependence on those quantities.

In order to have the initial velocities $\mathbf{v}_0$ uniformly distributed over a half sphere with $v_0={\rm const}$ in the velocity space, the values of the $\mu_0$ and $\phi_0$ are taken within a regular grid of $N_\mu \times N_\phi$ points: $\left\{ (\mu_{0;i},\phi_{0;j}) \right\}$, where $\mu_{0;i}=i/N_\mu$, $\phi_{0;j}=2\pi j/N_\phi$, with $i=1,...,N_\mu$, $j=0,...,N_\phi-1$. Typically, we used $N_\mu=N_\phi=1000$, which corresponds to a total number of particles $N_{\rm tot}=10^6$. Starting from those initial conditions, the motion equations (\ref{motioneq1}) and (\ref{motioneq2b}) are numerically integrated in time for all particles, until each particle has permanently left the SB (see above). The final value of the pitch angle cosine (which depends on both $\mu_{0;i}$ and $\phi_{0;j}$) is  $\mu_{1;i,j}$ and the corresponding variation is $\Delta \mu_{i,j} = \mu_{1;i,j}-\mu_{0,i}$. We note that $0< \mu_{0;i} \le 1$, while $-1 \le \mu_{1;i,j} \le 1$; therefore the variation $\Delta \mu_{i,j}$ varies in the interval $-2 \le \Delta \mu_{i,j} < 1$. To simplify the notation, from now on we drop the indexes $i$ and $j$. We calculated the distribution of the pitch-angle cosine variations $f(\Delta \mu)$ and the distribution of the final pitch-angle cosine $g(\mu_1)$ for different choices of the energy of particles and of the parameters characterising the magnetic structure of the SB.

\subsection{Varying reversal depth}

   \begin{figure}
   \centering
    \includegraphics[width=\hsize]{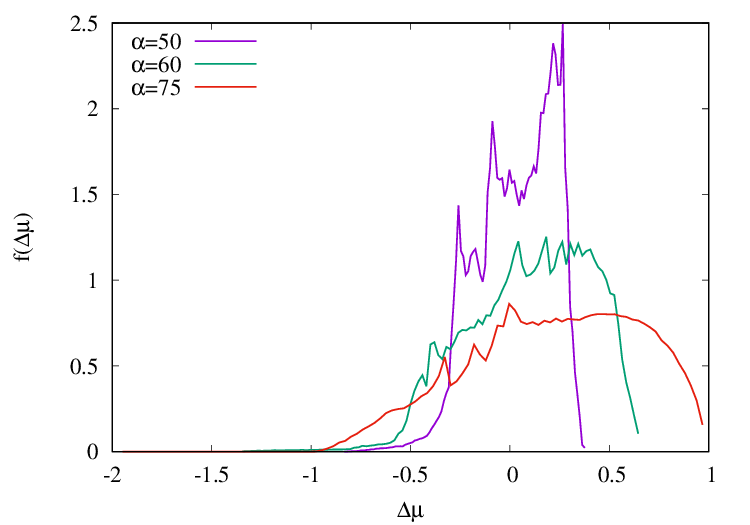}
    \includegraphics[width=\hsize]{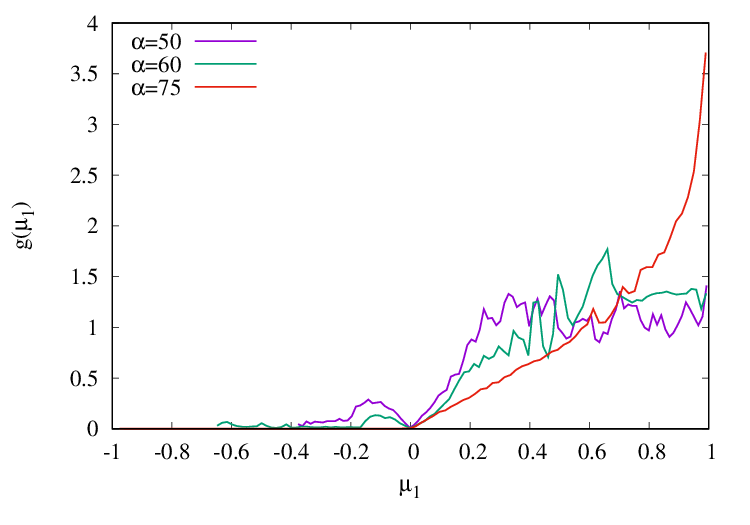}
   \caption{Distributions $f(\Delta \mu)$ of pitch-angle cosine variations (top panel) and $g(\mu_1)$ of final pitch-angle cosine (bottom panel) for $\beta=90^\circ$, $\alpha=50^\circ, 60^\circ, 75^\circ$, and particle energy $\mathcal{E}=1$ MeV.}
              \label{Fig:dmu_mu1_alpha50-75}%
    \end{figure}
We consider a case where all particles have energy $\mathcal{E}=1$ MeV, corresponding to $\rho_{\rm max}=1$  (see Fig. \ref{Fig:rho_vs_E}). In this case, particles trajectories have a Larmor radius of the order of the width of the two RDs. As shown in Paper I, in this condition there is a strong effect of the RD on the particle dynamics, which can lead to a relevant modification of the pitch angle. A similar behaviour is expected in the present case as well.

In Fig. \ref{Fig:dmu_mu1_alpha50-75} (top panel), the distribution $f(\Delta \mu)$ of pitch-angle cosine variations is plotted for $\beta=90^\circ$ and $\alpha=50^\circ, 60^\circ, 75^\circ$, that is, for a reversal depth going from mild to very pronounced.  The pitch-angle variation that a particle experiences depends on the initial condition $\left\{ \mu_0, \phi_0 \right\}$ of the given particle. 
%JG: I didnt understand this sentence very well, so replaced it with what I think you mean. If I misunderstood, just remove my version, and replace it.
%This gives origin to scattering in pitch angle. 
This gives the change from the initial to final pitch angle.
From Fig. \ref{Fig:dmu_mu1_alpha50-75}, we see that the width of the $f(\Delta \mu)$ distribution increases with increasing depth of the magnetic reversal, corresponding to larger variations in pitch angle. However, such a diffusion process is not symmetric; indeed, positive values of $\Delta \mu$ prevail over negative values and this asymmetry tends to increase for increasing obliquity angle $\alpha$.

In the bottom panel of Fig. \ref{Fig:dmu_mu1_alpha50-75}, the distribution $g(\mu_1)$ of the final pitch angle cosine is plotted for the same parameter values as in the top panel. Though initial values $\mu_0$ are uniformly distributed in the interval $] 0, 1 ]$, in the distribution $g(\mu_1)$ there is a prevalence of values close to $\mu_1 \sim 1$, corresponding to particles with $v_{||} \gg v_\perp$. This tendency becomes more evident with increasing obliquity angle $\alpha$; for $\alpha=75^\circ$ (deep magnetic reversal) the particle velocity distribution strongly focuses in the direction parallel to $\mathbf{B}$. A small population of particles with $\mu_1 < 0$ is also present. Those particles have a negative final $v_{||}$, and are therefore reflected back by the SB. The number of reflected particles decreases with increasing $\alpha$. 

In summary, when the particle Larmor radius is of the order of the RD width ($\mathcal{E}=1$ MeV), increasing the reversal depth leads to both larger pitch-angle diffusion and to velocity focusing in the parallel direction, along with a lower percentage of reflected particles.

\subsection{Varying particle energy}
%-------------------------------------------------------------
%                                             Simple A&A Table
%-------------------------------------------------------------
%
\begin{table}
\caption{Proton Larmor radius}             % title of Table
\label{table:1}      % is used to refer this table in the text
\centering                          % used for centering table
\begin{tabular}{c c c c}        % centered columns (4 columns)
\hline\hline                 % inserts double horizontal lines
$\mathcal{E}$ (MeV) & $\rho_{\rm max}$ (km) & $\rho_{\rm max}/\Delta x$ & $\rho_{\rm max}/(2x_c)$ \\    % table heading 
\hline                        % inserts single horizontal line
   $0.1$ & $3.05 \times 10^3$ & $0.320$ & $3.20 \times 10^{-2}$ \\      % inserting body of the table
   $0.2$ & $4.31 \times 10^3$ & $0.452$ & $4.52 \times 10^{-2}$ \\
   $0.3$ & $5.28 \times 10^3$ & $0.554$ & $5.54 \times 10^{-2}$ \\
   $0.5$ & $6.81 \times 10^3$ & $0.716$ & $7.16 \times 10^{-2}$ \\
   $1$ & $9.64 \times 10^3$ & $1.01$ & $0.101$ \\
   $10$ & $3.05 \times 10^4$ & $3.21$ & $0.321$ \\
   $10^2$ & $9.89 \times 10^4$ & $10.4$ & $1.04$ \\
   $10^3$ & $3.77 \times 10^5$ & $39.6$ & $3.96$ \\
   $3\times 10^3$ & $8.51 \times 10^5$ & $89.3$ & $8.93$ \\   
   $10^4$ & $2.42 \times 10^6$ & $255$ & $25.5$ \\   
   \hline                                   %inserts single line
\end{tabular}
%\caption{Values of proton Larmor radius $\rho_{\rm max}$ in physical units and normalized to the RD width $\Delta x$ and to the SB width $2x_c=10$, for the values of the proton energy $\mathcal{E}$ considered in Fig. \ref{Fig:rho_vs_E}}
\end{table}
%

\begin{figure}
   \centering
    \includegraphics[width=\hsize]{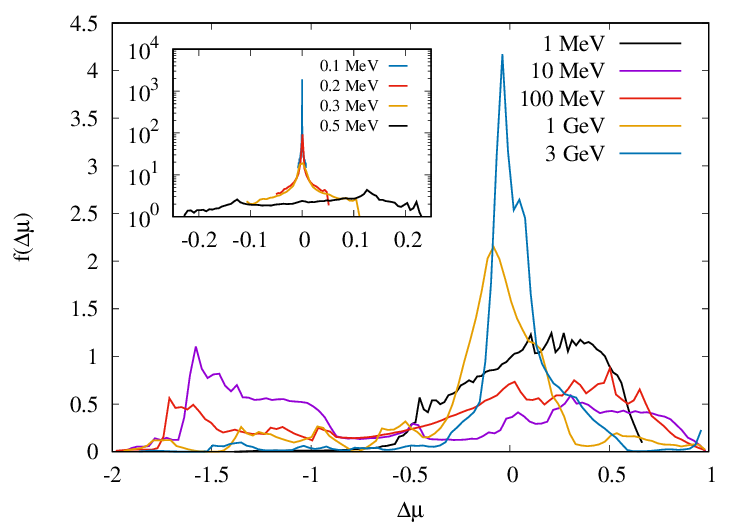}
    \includegraphics[width=\hsize]{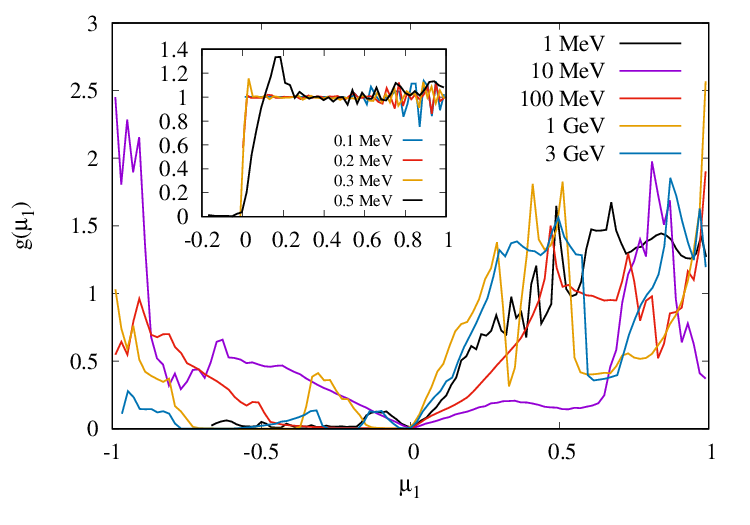}
   \caption{Distributions $f(\Delta \mu)$ of pitch-angle cosine variations (top panel) and $g(\mu_1)$ of final pitch-angle cosine (bottom panel) calculated for various values of the particle energy $\mathcal{E}$ in the  range $0.1$ MeV $\le \mathcal{E}\le 3\times 10^3$ MeV. The cases $\mathcal{E}=0.1,0.2,0.3,0.5$ MeV are shown in the insets. Both panels refer to a magnetic structure where $\beta=90^\circ$ and $\alpha=60^\circ$.
   }
              \label{Fig:dmu_mu1_energy}%
    \end{figure}

\begin{figure}
   \centering
    \includegraphics[width=\hsize]{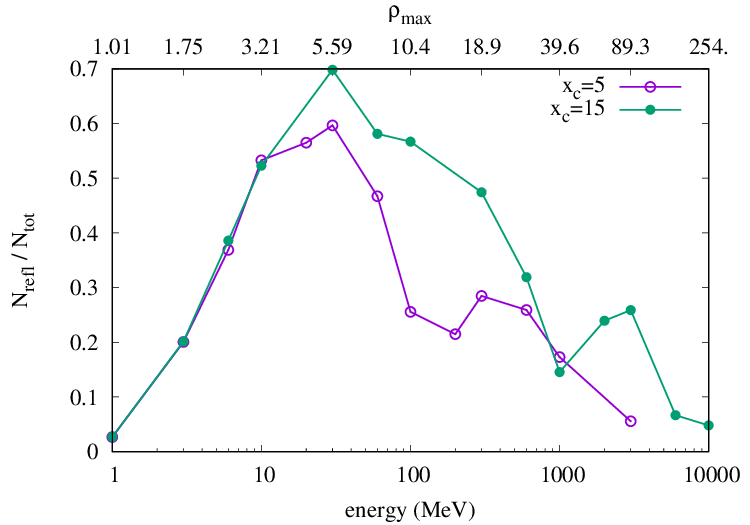}
   \caption{Fraction $N_{\rm refl}/N_{\rm tot}$ of reflected particles plotted as a function of the particle energy $\mathcal{E}$ for two values of SB width: $2x_c=10$ and $2x_c=30$. Both curves correspond to $\beta=90^\circ$, $\alpha=60^\circ$, and $\Delta x=1$. The values of $\rho_{\rm max}$ are indicated in the top horizontal axis for reference. 
   %{\bf In questa figura, ci starebbero bene dei simboli nei data points, ad es. dei pallini}
   }
              \label{Fig:reflect}%
    \end{figure}

%Different values of the particle energy $\mathcal{E}$ correspond to different Larmor radii. 
The Larmor radius $\rho_{\rm max}$ can be compared with the typical lengths in the magnetic structures, namely the RD width $\Delta x=1$ and the SB width $2x_c$. In order to study different regimes, we considered an energy interval going from $\mathcal{E}=0.1$  MeV, corresponding to $\rho_{\rm max}\simeq 0.32 < \Delta x$, up to $\mathcal{E}=10^4$  MeV, corresponding to $\rho_{\rm max}\simeq 255 \gg 2x_c$ 
(Fig. \ref{Fig:rho_vs_E}). Values of $\rho_{\rm max}$ corresponding to values of the particle energy $\mathcal{E}$ used in Figs. \ref{Fig:dmu_mu1_energy} and \ref{Fig:reflect} are listed in Table \ref{table:1} in physical units, and are also given as values normalised to the RD width $\Delta x$ and to the SB width $2x_c$ in the case $x_c=5$.

The top panel of Fig. \ref{Fig:dmu_mu1_energy}  shows the distributions $f(\Delta \mu)$ of pitch-angle cosine variations calculated for various values of the particle energy $\mathcal{E}$ ranging in the interval $0.1$ MeV $\le \mathcal{E}\le 3\times 10^3$ MeV. The cases $\mathcal{E}=0.1, 0.2, 0.3, 0.5$ MeV are shown in the inset. All distributions are calculated for $\alpha=75^\circ$ and $\beta=90^\circ$, corresponding to a relatively deep magnetic reversal (Fig. \ref{Fig:Br}). 

In the case where $\mathcal{E}=0.1$ MeV, the distribution $f(\Delta \mu)$ is strongly peaked around the value $\Delta \mu = 0$, indicating that the pitch-angle of each particle is almost unchanged when crossing the SB: $\mu_1 \simeq \mu_0$. This behaviour is a consequence of the conservation of the magnetic moment, $\mu_B = m_p v_\perp^2/(2B)\simeq {\rm const}$, which is satisfied when the Larmor radius is much smaller than the RD width. As the magnetic field intensity $B$ is uniform  in
our configuration, the above condition implies $v_\perp \simeq {\rm const}$. Moreover, as $v={\rm const,}$ it follows that also $v_{||}\simeq {\rm const}$. Therefore, $\mu=v_{||}/v$  remains approximately constant in time.

%% JG Comment about this next paragraph. It is intereting that there is such a big change from E=0.1MeV to E=1.0MeV.It might be interesting to plot two intermediate energies E=0.2MeV and E=0.5MeV to see the transition from the case in which magnetic moment is conserved and when it is not 
With increasing particle energy $\mathcal{E}$ from $0.1$ MeV to $0.5$  MeV, the width of the distribution $f(\Delta \mu)$ gradually increases, indicating that magnetic moment conservation is progressively lost. Further increasing $\mathcal{E,}$ the distribution $f(\Delta \mu)$ broadens until it covers almost all of the allowed interval $-2 \le \Delta \mu \le 1$ at $\mathcal{E}=10$ MeV. This feature is found up to energies $\mathcal{E}\sim 100$ MeV, corresponding to $\rho_{\rm max}\sim 10$. Therefore, when the Larmor radius varies between the width $\Delta x$ of the RDs up to the width $2x_c$ of the SB, particles undergo a very significant pitch-angle scattering. 

Further increasing the energy $\mathcal{E}$ leads to the opposite behaviour: the distributions $f(\Delta \mu)$ at $\mathcal{E}=10^3$ MeV and $\mathcal{E}=3\times 10^3$ MeV become increasingly peaked around $\Delta \mu=0$. This regime corresponds to $\rho_{\rm max} \gg 2x_0$. Therefore, when the Larmor radius is much larger than the SB width, particles `jump' across the SB, remaining much less affected than at lower energies. We conclude that the regime of maximum pitch-angle scattering corresponds to $\rho_{\rm max}$ in between the RD width and the SB width.%, in analogy to what has been observed in shock physics \citep{}.

The distribution $g(\mu_1)$ of the final pitch-angle cosine is shown in the bottom panel of Fig. \ref{Fig:dmu_mu1_energy} for different values of the particle energy in the range $0.1$ MeV $\le \mathcal{E}\le 3\times 10^3$ MeV. The cases $\mathcal{E}=0.1, 0.2, 0.3, 0.5$ MeV are shown in the inset. At low energies, the distribution $g(\mu_1)$ is very close to the initial pitch-angle cosine distribution, and is almost constant in the range $0 \lesssim \mu_1 \lesssim 1$. By increasing $\mathcal{E}$, $g(\mu_1)$ starts to depart from a constant profile. This is consistent with the behaviour of $f(\Delta \mu)$ shown in the inset of the upper panel.
For intermediate energies, a relevant population of reflected particles, characterised by $\mu_1 <0$ is present. This is particularly evident in the distribution at $\mathcal{E}=10$ MeV (purple curve, corresponding to $\rho_{\rm max}$ comprised between the RD and the SB widths) where, after crossing the SB, particles focus in two beams, respectively almost parallel and anti-parallel to the magnetic field direction. Reflected particles are less copious at lower and higher energies.

The abundance of particles reflected by the SB can be considered as one of the possible measures of the effectiveness of the SB in affecting particle propagation. Figure \ref{Fig:reflect} shows the fraction $N_{\rm refl}/N_{\rm tot}$ of reflected particles over the total particle number $N_{\rm tot}$ in the energy range 1 MeV $\le \mathcal{E} \le 10^4$ MeV and for two values of the SB width: $2x_c=10$ and $2x_c=30$. Both curves correspond to $\beta=90^\circ$, $\alpha=60^\circ$, and $\Delta x=1$. As expected, we see that the fraction of reflected particles is more relevant at intermediate energies, that is, for Larmor radii comprised between the RD width and the SB width. The fraction $N_{\rm refl}/N_{\rm tot}$ reaches high values (around $60 \% - 70\%$) at energies $\mathcal{E}\sim 30$ MeV, corresponding to $\rho_{\rm max} \sim 5-6$. When increasing the SB width, the energy range where particle reflection is relevant extends towards higher energies, and the maximum fraction $N_{\rm refl}/N_{\rm tot}$ slightly increases. We also note that, at low energies ($\mathcal{E}\lesssim 10$ MeV), the two curves corresponding to different SB widths are superposed. This indicates that in the low-energy regime, particle dynamics is mainly regulated by their interaction with the two RDs. Instead, at higher energies, the effect of the entire SB width becomes more relevant and the two curves deviate from each other. It is interesting to notice that a secondary peak at high energies is visible in both curves in Fig. \ref{Fig:reflect}: we reserve the investigation of this feature to a future study.

 %  \begin{figure}
 %  \centering
 %   \includegraphics[width=\hsize]{w_theta_phi_paper.pdf}
 %  \caption{Two-dimensional maps of the final value $\mu_1$ of the pitch-angle cosine are plotted as functions of the initial values $\mu_0$ and $\phi_0$ for different values of the particle energy: $\mathcal{E}=1$ MeV (top panel), $\mathcal{E}=30$ MeV  (middle panel), $\mathcal{E}=1$ GeV  (bottom panel). The purple line is the contour separating regions where $w_1>0$ from regions where $w_1<0$. All plots are calculated for $\beta = 90^\circ$ and $\alpha = 60^\circ$. {\bf Ricordarsi di cambiare w in $\mu$ sull'asse delle ascisse}}
 %             \label{Fig:w1_vs_theta_phi}%
 %   \end{figure}

\subsection{Particle dynamics and chaos}
   \begin{figure*}
   \centering
    \includegraphics[width=\hsize]{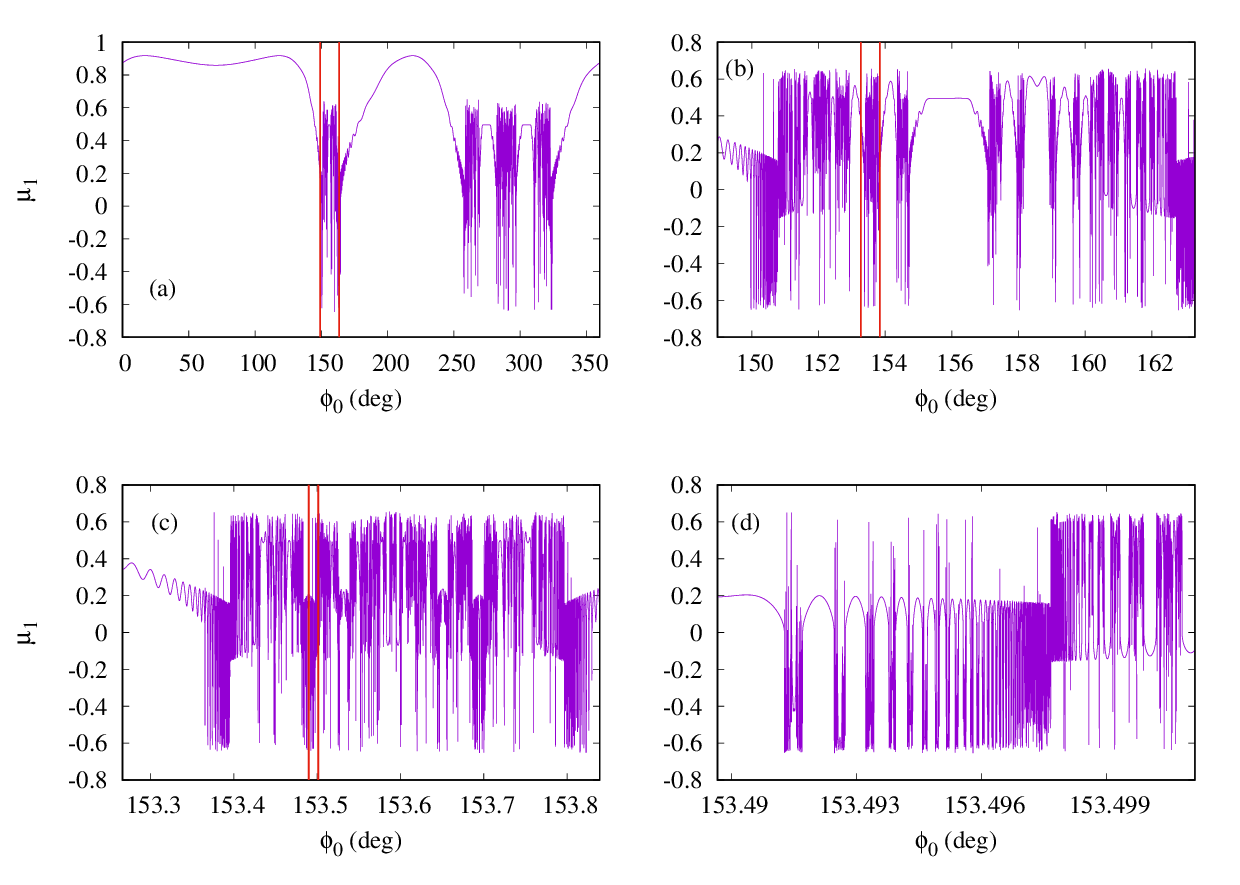}
   \caption{Final pitch-angle cosine $\mu_1$ as a function of the initial gyrophase $\phi_0$ for $\mu_0=0.5$ in the case $\beta=90^\circ$, $\alpha=60^\circ$, and $\mathcal{E}=1$ MeV (purple curves). Plot (a) refers to the whole range $0\le \phi_0 \le 360^\circ$, and plots (b) to (d) refer to increasingly small subintervals. Vertical red lines in plots (a) to (c) indicate the subinterval represented in the subsequent plot.}
              \label{Fig:multi}%
    \end{figure*}
The presence of a fine structure is apparent in the final pitch-angle cosine distribution $g(\mu_1)$ (Figs. \ref{Fig:dmu_mu1_alpha50-75} and \ref{Fig:dmu_mu1_energy}). We verified that those features do not change when increasing the number $N_{\rm tot}$ of particles and therefore cannot be ascribed to noise caused by insufficient counting statistics. Instead, they are related to a complex particle dynamics that generates structures and deterministic chaos in the phase space, as deeply discussed in Paper I. Indeed, in the SB model, we also observe significant variations in the final particle pitch angle values as tiny variations of the particle initial conditions (in $\mu_0$ and $\phi_0$) are considered.

To illustrate the chaotic behaviour, in Fig. \ref{Fig:multi} we plot the final pitch-angle cosine $\mu_1$ as a function of the initial gyrophase $\phi_0$ for a fixed value of the initial pitch-angle cosine, $\mu_0=0.5$, 
%(indicated by a vertical red line in Fig. \ref{Fig:w1_vs_theta_phi}, top panel) 
and for $\beta=90^\circ$, $\alpha=60^\circ$, and $\mathcal{E}=1$ MeV. In panel (a) the whole range $0\le \phi_0 \le 360^\circ$ is represented. For almost the whole interval, $\mu_1 >0$, which corresponds to forward-moving particles, and $\mu_1$ has a smooth dependence on the initial gyrophase $\phi_0$. However, some subranges are present where $\mu_1$ as a function of $\phi_0$ shows very fast variations and can take negative values, corresponding to back-reflected particles. A zoom onto one of these subranges is shown in Fig. \ref{Fig:multi}((b); indicated by two vertical red lines in panel (a)), where a structure similar to that in Fig. \ref{Fig:multi}(a) is visible at a smaller scale; namely, a succession of subdomains where $\mu_1(\phi_0)$ has either a smooth or a rapidly changing behaviour. Further progressive enlargements of such subdomains plotted in Figs. \ref{Fig:multi}(c) and \ref{Fig:multi}(d) display the same behaviour at increasingly small scale. 
%Two properties can be inferred from Fig. \ref{Fig:multi}: (i) Around regions of the $(\mu_0, \phi_0)$ plane where the final pitch-angle cosine $\mu_1$ is negative, $\mu_1$ is extremely sensitive to the initial condition, since very small variations of $\phi_0$ and $\mu_0$ lead to very different final values $\mu_1$; this kind of behavior is typical of a chaotic dynamics. (ii) The line in the $(\mu_0,\phi_0)$ plane separating regions where $\mu_1<0$ from those where $\mu_1>0$ appears to have self-similar properties and then it could be a fractal. 
Similar properties have been found in the dynamics of energetic particles propagating across a single RD (Paper I).

\begin{figure}
   \centering
    \includegraphics[width=\hsize]{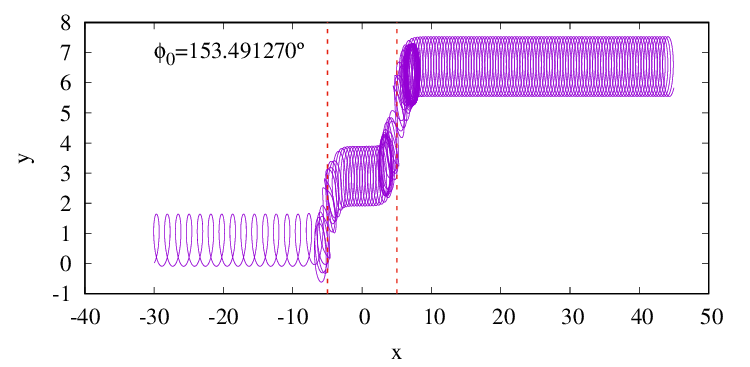}
    \includegraphics[width=\hsize]{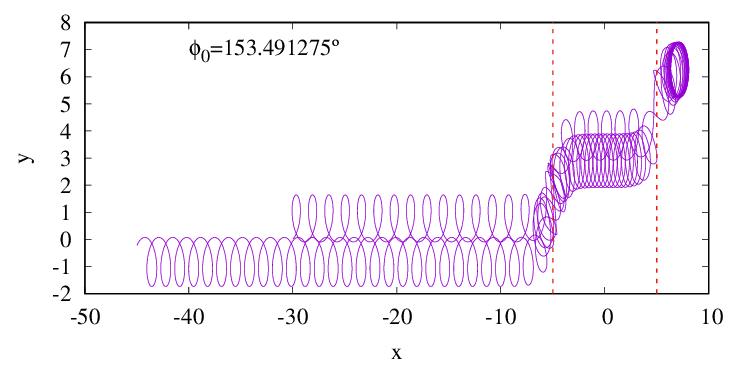}
    \includegraphics[width=\hsize]{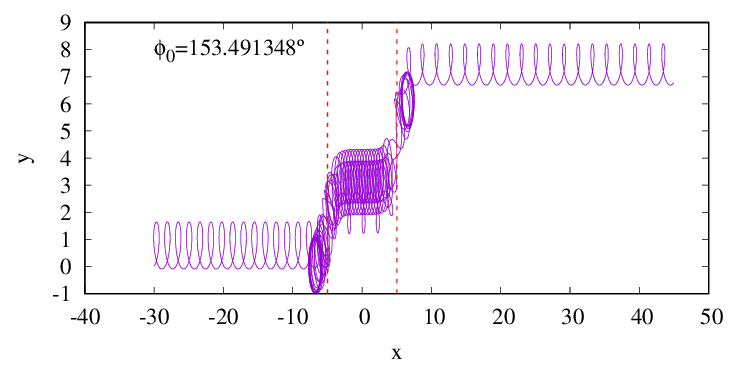}
   \caption{Trajectories of three particles (particle A: top panel; particle B: middle panel; particle C: bottom panel) projected onto the $xy$ plane. Each particle starts with a slightly different initial gyrophase $\phi_0$, the value of which is indicated in each panel, while the initial pitch-angle cosine is the same for the three particles ($\mu_0=0.5$). The values of the parameters $\beta$, $\alpha,$ and $\mathcal{E}$ are the same as in Fig. \ref{Fig:multi}. The positions of the two RDs are indicated by red dashed lines.}
              \label{Fig:traject}%
    \end{figure}

%% JG Comment/question: how sensitive is it to the numerical method? What is the accuracy of the numerical method? Does this matter? I think a reader might wonder the same thing, so it might be good to add a sentence or two about it.
To illustrate the extreme sensitivity of the particle dynamics to initial conditions in the chaotic regions, Fig. \ref{Fig:traject} reports the trajectories of three particles ---denoted A, B and C--- that are injected with $\mu_0=0.5$ and $\phi_{0A}=143.491270^\circ$, $\phi_{0B}=143.491275^\circ$ and $\phi_{0C}=143.491348^\circ$. Trajectories are projected onto the $xy$ plane. The positions of the two RDs are indicated by red dashed lines. The values of the other parameters are the same as in Fig. \ref{Fig:multi}. Though the initial conditions are very close to one another (the relative difference in the initial gyrophases is $\lesssim 5\times 10^{-7}$), the subsequent time evolution is completely different. In particular, particle A enters the SB crossing the first RD and reaches the second RD. It then remains trapped for a certain time inside the second RD, finally exiting from the opposite side. Particle trapping inside RDs was also demonstrated in Paper I. It can be noticed that the final pitch-angle of particle A is very different from the initial one. 
%this phenomenon is at the base of the pitch-angle scattering. 
Particle B, which has an initial gyrophase very close to that of particle A ($|\phi_{0B}-\phi_{0A}|/\phi_{0B}=3.5\times 10^{-8}$), initially behaves in a similar way to particle A, until it is trapped inside the second RD. However,  particle B subsequently exits the RD, moving back in the negative-$x$ direction; it then crosses the first RD  again, permanently exiting the SB. Therefore, particle B is classified as a reflected particle. The dynamics of particle C is even more complex: it crosses the first RD, reaches the second one, and is then reflected back towards the first. Here, it experiences a further reflection in the positive $x$ direction. Finally, it crosses the second RD leaving the SB. Therefore, though particle C is not classified as a reflected particle ($\mu_1>0$), its dynamics includes multiple reflections inside the SB. These different particle time evolutions, that is, reflected, trapped, and transmitted, may be compared with those reported in Fig. 1 of \citet{Moraal13}, and the values of the initial conditions of the trajectories in Fig. \ref{Fig:traject} show that it is very difficult to devise a priori the behaviour of a particle interacting with a SB.

Finally, we notice that the presence of chaos in specific regions of the $(\mu_0,\phi_0)$ plane implies that in those regions the calculation of single-particle evolution becomes sensitive to details of the numerical method. For instance, the trajectories shown in Fig. \ref{Fig:traject} change when changing the time step. This is an unavoidable aspect of chaotic dynamics. However, we verified that the distributions $f(\Delta \mu)$ and $g(\mu_1)$ shown in Figs.  \ref{Fig:dmu_mu1_alpha50-75} and \ref{Fig:dmu_mu1_energy} remain unchanged when reducing the time step. Therefore, from a statistical point of view, our results are not sensitive to the numerical method.

\section{Conclusions}
We studied the dynamics of energetic charged particles in a non-uniform magnetic field $\mathbf{B}$, which represents a simplified model of a magnetic SB. Recent measurements performed by spacecraft, such as PSP and SolO, revealed that SBs are commonly present in the inner heliosphere. Those structures can affect the transport of energetic particles. Our model emphasises relevant features characterising $\mathbf{B}$ in SBs, namely the presence of abrupt $\mathbf{B}$ rotations (RDs) limiting the SB and a nearly uniform magnetic field intensity $|\mathbf{B}|$. The model depends on some parameters, in particular the obliquity angle $\alpha$ and the rotation angle $\beta$; a magnetic field reversal from outside to inside the magnetic structure is obtained for large values of $\alpha$ and for $\beta \sim 90^\circ$. %Other parameters are the SB width $2x_c$ and the RD thickness $\Delta x$. 
The evolution of mono-energetic populations of particles impinging on the SB has been studied numerically by employing a symplectic integrator for the solution of relativistic motion equations. Energy conservation implies a possible exchange between parallel and perpendicular energy, which generates pitch-angle scattering. 

Results have been characterised by 
%\LEt{***The intended meaning here is unclear; please consider rewording and/or expanding.***}
calculating the distributions of the variations of the pitch-angle cosine $f(\Delta \mu)$ and of the final pitch-angle cosine $g(\mu_1)$ for uniformly distributed initial values $\mu_0$. In Paper I, where the effect of a single RD is examined, we found a relevant pitch-angle scattering when the particle Larmor radius $\rho$ is comparable to the RD thickness $\Delta x$. Similarly, in the present model, we find a relevant pitch-angle scattering for values of $\rho$ comprised within the interval $\rho \sim \Delta x$ up to $\rho \gtrsim 2x_c$, with $2x_c$ being the SB width. Moreover, for $\rho \ll \Delta x,$ the conservation of both magnetic moment $\mu_B$ and energy $\mathcal{E}$ implies near constancy of $\mu$. On the other hand, when $\rho \gg 2 x_c$, our model indicates a low level of pitch-angle scattering; when the particle Larmor radius is much larger than the SB width, the associated magnetic field inhomogeneity does not modify the particle pitch angle in a significant way. The latter result is analogous to the observation of adiabatic motion for energetic particles at oblique and quasi-perpendicular shock waves when particle gyroradii are much larger than the shock thickness, with a very small difference between the magnetic moment of a particle before and after a single shock encounter \citep{Whipple86,Decker88}.

For the parameter values used in the present model, the most relevant effect of the SB on the particle dynamics is in the energy interval $1$ MeV $\lesssim \mathcal{E} \lesssim $ few GeV. With this range, wide distributions $f(\Delta \mu)$ of the variation of the pitch-angle cosine have been found, covering the whole interval $-2 \lesssim \Delta \mu \lesssim 1$, mostly for energies $\mathcal{E} \sim 10-100$ MeV. Moreover, the width of $f(\Delta \mu)$ tends to increase with increasing depth of the magnetic reversal. In the same energy interval, we find a population of reflected particles that constitute up to $\simeq 70\%$ of the total particle population. 

Therefore, our results indicate the existence of a regime of intermediate energy $\mathcal{E}$ (or Larmor radii) where the effects of the SB on the particle dynamics is particularly relevant. For lower or higher energies, particles are much less affected by the SB. 
This is partly in agreement with the findings of \citet{Bandyopadhyay21} from their analysis of high-energy particles as detected by the EPI-Lo instrument on board PSP, where the ion flux was not changing from antisunward to sunward within the SB for particles with gyroradii comparable with the SB size. In our model, large pitch-angle changes tend to be reduced for particles with gyroradii larger than the SB width. However, in our model, we checked the pitch-angle variations upstream of the discontinuity, not through it, and the magnetic field value is different from the dataset analysed in \citet{Bandyopadhyay21}; comparisons should therefore be made with caution.

Of course, the values of the energy $\mathcal{E}$ quoted above can vary when the values of the parameters characterising the model are changed. In particular, we considered a SB width of $2x_c=10-30$ (in normalised units), which corresponds to a crossing time $t_c=280-840$ s (assuming $\ell=9.52 \times 10^3$ km and $v_{SW}=340$ km $s^{-1}$). Indeed, an analysis of SB duration has shown that the SB width can vary by about two orders of magnitude, in the interval $t_c \sim 10^2-10^4$ s \citep{Pecora22}. Therefore, we expect that the regime where the effect of a SB on particles is relevant can be found at different energies according to both the SB width and the RD thickness.

The presence of chaos is another feature of the dynamics of particles propagating across a RD (see Paper I). This property has also been found in our SB model. In particular, there are regions in the space of initial condition, where very small variations in the particle initial condition lead to completely different trajectories. In those regions, we verified that particles starting with the same pitch angle go beyond the SB, are reflected back, or undergo multiple reflections between the RDs as a consequence of variations in the initial gyrophase by an amount as small as $\sim 5\times 10^{-5} \%$. Such extreme sensitivity to initial conditions is one of the features that characterises a chaotic dynamics
%\LEt{***Please check that I have retained your intended meaning.***}
. Chaotic regions are mostly concentrated at large initial pitch angles, but they can also be found at lower pitch angles, especially when the fraction of reflected particles $N_{\rm refl}/N_{\rm tot}$ is large. Therefore, values of energy (or Larmor radius) giving rise to a more relevant effect of SB on particles also correspond to larger chaotic regions in the initial condition space. In this connection, we recall that the presence of chaotic scattering regions in phase space can influence not only particle propagation but also processes such as magnetic reconnection \citep{Buchner87,Buchner89}.

% JG comment: Just a side comment. This is interesting to me. I wonder if there is a way to quantitatively determine the difference between scattering via interaction with switchbacks, vs. scattering via regular turbulence (like quasi-linear theory). It certainly must depend on the distribution of swtichbacks.
We note that the pitch-angle scattering resulting from the interaction between particles and a SB described here is different from the perturbative, small-angle scattering considered in diffusion theories, because large-angle scattering is often prevailing, as shown in Fig. \ref{Fig:dmu_mu1_energy}. Indeed, the probability distribution functions of pitch-angle variations are far from bell-shaped functions. Also, the particle reflection process is different from magnetic mirroring, because the amplitude of the magnetic field is constant and the fraction of reflected particles also depends on energy, as shown in Fig. \ref{Fig:reflect}

Our results show that the interaction of energetic particles with magnetic SB can have a number of consequences. For instance, if a solar energetic particle (SEP) event is impinging on a SB, which is ahead of the energetic particles, that is, is farther away from the Sun, a large number of these particles can be scattered back towards the Sun, meaning that the intensity of SEPs beyond the SB will be decreased, giving rise to possible dropouts in the energetic particle fluxes. We note that, according to Figure \ref{Fig:reflect}, the energy range in which a dropout is expected can be predicted if the SB width is obtained from the measured SB duration and the solar wind speed. At the same time, transmitted particles may become more field aligned, as shown by the distribution of final pitch-angle cosines in Figures \ref{Fig:dmu_mu1_alpha50-75} and \ref{Fig:dmu_mu1_energy}. It would be interesting if this could be checked in spacecraft measurements, looking at the particle (ions and electrons) fluxes sampled at different pitch-angles.
We believe that with multi-spacecraft observations it should be possible to check these properties, if simultaneous SEP measurements by magnetically connected spacecraft on both sides of a SB are available. 

Another consequence is related to the fact that SBs increase pitch-angle scattering for the range of energies outlined above, and in particular they increase large pitch-angle scattering: this can have an influence on the processes of Fermi acceleration, both first and second order, because, as we show here, a SB can reflect particles very efficiently. In some sense, SBs can act as magnetic mirrors, even if the magnetic field magnitude is constant. We propose that these effects should be taken into account when studying energetic particle propagation and acceleration. 

% JG Comment about this. Just wanted to alert you to an analogy. I did my PhD thesis on conservation of magnetic moment at shocks, which is analagous to this situation. A shock has a thickness that is much less than the gyroradius of the particles of interest. Yet, I found that the magnetic moment of the particle caclculated well before the particle reached the shock was very close to the same magnetic moment well after the shock. That is, even though the magnetic moment changes dramatically at each shock crossing, the initial and final mag. moments were very nearly the same. This suggests there is some invariant of the particle motion in the limit when the structure thickness is very small compared to the particle gyroradii. Here are three relevant references. In addition to these, I believe David Burgess's 1987 paper on shock drift acceleration also looked at this.
%
% Whipple, E. C., T. G. Northrup, and T. J. Birmingham, Adiabatic theory in regions of strong field gradients, J. Geophys. Res., 91, 4149, 1986.
% Decker, R. B., Computer modeling of test-particle acceleration at oblique shocks, Space Sci. Rev., 48, 195, 1988.
% Giacalone, J., Proton Acceleration in Structured Collisionless Shock Waves, PhDT, University of Kansas, 1991.
% 

\begin{acknowledgements}
      S. P. and G. Z. acknowledge support from the Italian Space Agency and the National Institute of Astrophysics, in the framework of the CAESAR (Comprehensive spAce wEather Studies for the ASPIS prototype Realization) project, through the ASI-INAF n. 2020-35-HH.0 agreement for the development of the ASPIS (ASI SPace weather InfraStructure) prototype of scientific data centre for Space Weather. F.M. and G.Z. acknowledge support from the Italian Space Agency through the HENON project and the MUSE project.     
\end{acknowledgements}


\begin{thebibliography}{}

\bibitem[Amato(2014)]{Amato14}
Amato, E. 2014, Int. J. Mod. Phys. D 23, 1430013

\bibitem[Artemyev et al.(2020)]{Artemyev20}
Artemyev, A. V., Neishtadt, A. I., Vasiliev,  et al. 2020, PRE, 102, 033201

\bibitem[Bale et al.(2019)]{Bale19}
Bale, S. D., et al. 2019, Nature, 576, 237--242

\bibitem[Bale et al.(2021)]{Bale21}
Bale, S. D., Horbury, T. S., Velli, M., et al. 2021, ApJ, 923, 174

\bibitem[Bandyopadhyay et al.(2021)]{Bandyopadhyay21}
Bandyopadhyay, R., et al. 2021, Astron. Astrophys., 650, L4

\bibitem[Belcher \& Davis(1971)]{Belcher71}
Belcher, J. W. \& Davis, L. 1971, JGR, 76, 3534 

\bibitem[Borovsky(2010)]{Borovsky10}
Borovsky, J. E. 2010, PRL, 105, 111102

\bibitem[Borovsky(2016)]{Borovsky16}
Borovsky, J. E. 2016, JGRA, 121, 5055

\bibitem[Bruno \& Carbone(2013)]{Bruno13}
Bruno, R. \& Carbone, V. 2013, Liv. Rev. Sol. Phys. 10, 2

\bibitem[Burlaga(1969)]{Burlaga69a}
Burlaga, L. F. 1969a, Solar Phys. 7, 54 

\bibitem[Burlaga \& Ness(1969)]{Burlaga69b}
Burlaga, L. F. \& Ness, N. F. 1969, Solar Phys. 9, 467 

\bibitem[Buechner \& Zelenyi(1987)]{Buchner87}
Buechner, J., \& Zelenyi, L.M. 1987, J. Geophys. Res. 92, 13456

\bibitem[Buechner \& Zelenyi(1989)]{Buchner89}
Buechner, J., \& Zelenyi, L.M. 1989, J. Geophys. Res. 94, 11821

\bibitem[Crooker et al.(1989)]{Crooker89}
Crooker, N. U., Gosling, J. T., Bothmer, et al. 1999, Space Sci. Rev. 89, 179

\bibitem[Decker(1988)]{Decker88}
Decker, R. B. 1988, Space Sci. Rev., 48, 195

\bibitem[Dudok de Wit et al.(2020)]{Dudok20}
Dudok de Wit, T., Krasnoselskikh, V. V., Bale, S. D., et al. 2020, ApJS, 246, 39

\bibitem[Fedorov et al.(2021)]{Fedorov21}
Fedorov, A., et al. 2021, A\&A, 656, A40

\bibitem[Fisk \& Kasper(2020)]{Fisk20}
Fisk, L. A., \& Kasper, J. C. 2020, ApJL, 894, L4

\bibitem[Florinski et al.(2003)]{Florinski03}
Florinski, V., Zank, G. P., \& Pogorelov, N. V., 2003, JGR 108, A6

\bibitem[Giacalone \& Jokipii(2001)]{Giacalone01}
Giacalone, J. \& Jokipii, J. R., 2001, Adv. Space Res. 27, 461

\bibitem[Giacalone(2013)]{Giacalone13}
Giacalone, J. 2013, Space Sci. Rev. 176, 73

\bibitem[Greco \& Perri(2014)]{Greco14}
Greco, A. \& Perri,  S. 2014, ApJ, 784, 163

\bibitem[Greco et al.(2016)]{Greco16}
Greco, A., Perri, S., Servidio, S., Yordanova, E. \& Veltri, P. 2016, ApJL, 823, L39 

\bibitem[Horbury et al.(2001)]{Horbury01}
Horbury, T., Burgess, D., Fr\"anz, M. J. \& Owen, C. J. 2001, GRL, 28, 677 

\bibitem[Horbury et al.(2018)]{Horbury18}
Horbury, T. S., Matteini, L., \& Stansby, D. 2018, MNRAS, 478, 1980

\bibitem[Horbury et al.(2020)]{Horbury20}
Horbury, T. S., Woolley, T., Laker, R., et al. 2020, ApJS, 246, 45

\bibitem[Hussein \& Shalchi(2016)]{Hussein16}
Hussein, M.  \& Shalchi, A. 2016, ApJ, 817, 136

\bibitem[Jokipii(1966)]{Jokipii66}
Jokipii, J. R. 1966, ApJ, 146, 480

\bibitem[Kasper et al.(2019)]{Kasper19}
Kasper, J. C. et al. 2019, Nature 576, 228

\bibitem[Knetter et al.(2003)]{Knetter03}
Knetter, T., Neubauer, F. M., Horbury, T. \& Balogh, A. 2003, Adv. Space Res., 32, 543

\bibitem[Knetter et al.(2004)]{Knetter04}
Knetter, T., Neubauer, F. M., Horbury, T. \& Balogh, A. 2004, JGR, 109, A06102

\bibitem[Laker et al.(2021)]{Laker21}
Laker, R., Horbury, T. S., Bale, S. D., et al. 2021, A\&A, 650, A1

\bibitem[Landi et al.(2006)]{Landi06}
Landi, S., Hellinger, P., \& Velli, M. 2006, GeoRL, 33, L14101

\bibitem[Lee \& Fisk(1982)]{Lee82}
Lee, M. A. \& Fisk, L. A. 1982, Space Sci. Rev. 32, 205

\bibitem[Lee et al.(2012)]{Lee12}
Lee, M. A., Mewaldt, R. A. \& Giacalone, J. 2012, Space Sci. Rev., 173, 247

\bibitem[McComas et al.(2016)]{McComas16}
McComas, D. J., Alexander, N., Angold, N., et al. 2016, Space Sci. Rev., 204, 187

\bibitem[Magyar et al.(2021a)]{Magyar21a}
Magyar, N., Utz, D., Erdélyi, R., \& Nakariakov, V. M. 2021a, ApJ, 911, 75

\bibitem[Magyar et al.(2021b)]{Magyar21b}
Magyar, N., Utz, D., Erdélyi, R., \& Nakariakov, V. M. 2021b, ApJ, 914, 8

\bibitem[Malara et al.(2021)]{Malara21}
Malara, F., Perri, S. \& Zimbardo, G. 2021, PRE, 104, 025208, Paper I

\bibitem[Mariani et al.(1983)]{Mariani83}
Mariani, F., Bavassano, B. \& Villante, U. 1983, Solar Phys., 83, 349 

\bibitem[Martin et al.(1973)]{Martin73}
Martin, R. N., Belcher, J. W. \& Lazarus, A. J. 1973, JGR, 78, 3653 

\bibitem[Matthaeus et al.(2003)]{Matthaeus03}
Matthaeus, W. H., Qin, G., Bieber, J. W. \& Zank, G. P. 2003, ApJL, 590, L53

\bibitem[McCracken \& Ness(1966)]{McCracken66}
McCracken, K., \& Ness, N. 1966, JGR, 71, 3315

\bibitem[McManus et al.(2020)]{McManus20}
McManus, M. D., Bowen, T. A., Mallet, A., et al. 2020, ApJS, 246, 67

\bibitem[Moraal(2013)]{Moraal13}
Moraal, H. 2013, Space Sci. Rev. 176, 299

\bibitem[Mozer et al.(2020)]{Mozer20}
Mozer, F. S., Agapitov, O. V., Bale, S. D., et al. 2020, ApJS, 246, 68

\bibitem[Mozer et al.(2021)]{Mozer21}
Mozer, F. S., Bale, S. D., Bonnell, J. W., et al. 2021, ApJ, 919, 60

\bibitem[Neugebauer(1989)]{Neugebauer89}
Neugebauer, M. 1989, GRL, 16, 1261 

\bibitem[Neugebauer \& Goldstein(2013)]{Neugebauer13}
Neugebauer, M., \& Goldstein, B. E. 2013, in AIP Conf. Proc., 1539, SOLAR WIND 13: Proc. of the Thirteenth International Solar Wind Conference
(New York: AIP), 46

\bibitem[Parizot et al.(2006)]{Parizot06}
Parizot, E., Marcowith, A., Ballet, J., \& Gallant, Y. A. 2006, A\&A 453, 387

\bibitem[Pecora et al.(2022)]{Pecora22}
Pecora, F., Matthaeus, W. H., Primavera, et al. 2022, ApJL, 929:L10

\bibitem[Perri et al.(2012)]{Perri12}
Perri, S., Goldstein, M. L., Dorelli, J. C. \& Sahraoui, F. 2012, PRL, 109, 191101 

\bibitem[Perri et al.(2017)]{Perri17}
Perri, S., Servidio, S., Vaivads,  A. \& Valentini, F. 2017, Astrophys. J. Suppl. Ser. 231, 4

\bibitem[Perrone et al.(2020)]{Perrone20}
Perrone, D., Bruno, R., D’Amicis, R., et al. 2020, ApJ, 905, 142

\bibitem[Phan et al.(2020)]{Phan20}
Phan, T. D., Bale, S. D., Eastwood, et al. 2020, ApJS, 246, 34 

\bibitem[Pommois et al.(2005)]{Pommois05}
Pommois, P., Zimbardo,  G. \& Veltri, P. 2005, Adv. Space Res. 35, 647

\bibitem[Pucci et al.(2016)]{Pucci16}
Pucci, F., Malara, F., Perri, S., et al. 2016, MNRAS, 459, 3395

\bibitem[Roberts(2012)]{Roberts12}
Roberts, D. A. 2012, PRL, 109, 231102 

\bibitem[Ruffolo et al.(2020)]{Ruffolo20}
Ruffolo, D., Matthaeus, W. H., Chhiber, R., et al. 2020, ApJ, 902, 94

\bibitem[Shalchi(2009)]{Shalchi09}
Shalchi, A. 2009, Nonlinear Cosmic Ray Diffusion Theories, Astrophysics and Space Science Library, Volume 362. ISBN 978-3-642-00308-0. S pringer-Verlag Berlin Heidelberg

\bibitem[Schwadron \& McComas(2021)]{Schwadron21}
Schwadron, N., \& McComas, D. 2021, ApJ, 909, 95

\bibitem[Smith(1973)]{Smith73}
Smith, E. J. 1973, JGR, 78, 2054 

\bibitem[Squire et al.(2020)]{Squire20}
Squire, J., Chandran, B. D. G., \& Meyrand, R. 2020, ApJL, 891, L2

\bibitem[Soding et al.(2001)]{Soding01}
Soding, A., Neubauer, F. M., Tsurutani, B., Ness, N. F. \& Lepping, R. P. 2001, Ann. Geophys. 19, 667 

\bibitem[Telloni et al.(2022)]{Telloni22}
Telloni, D., Zank, G. P., Stangalini, M. et al. 2022, ApJL, 936:L25

\bibitem[Tenerani et al.(2020)]{Tenerani20}
Tenerani, A., Velli, M., Matteini, L., et al. 2020, ApJS, 246, 32

\bibitem[Tenerani et al.(2021)]{Tenerani21}
Tenerani, A., Sioulas, N., Matteini, L., et al. 2021, ApJL, 919, L31

\bibitem[Tessein et al.(2015)]{Tessein15}
Tessein, J. A., Ruffolo, D., Matthaeus, W. H., et al. 2015, ApJ, 812, 68

\bibitem[Tsurutani \& Smith(1979)]{Tsurutani79}
Tsurutani, B. T. \& Smith,  E. J. 1979, JGR, 84, 2773

\bibitem[Valentini et al.(2019)]{Valentini19}
Valentini, F., Malara, F., Sorriso-Valvo,  L., Bruno, R. \& Primavera, L. 2019, ApJL, 881, L5 

\bibitem[Veltri \& Mangeney(1999)]{Veltri99}
Veltri, P. \& Mangeney, A. 1999, in Solar Wind Nine, American Institute of Physics Conference Series, Vol. 471, Scaling laws and intermittent structures in solar wind MHD turbulence, ed.s S. R. Habbal, R. Esser, J. V. Hollweg, \& P. A. Isenberg, pp. 543–546.

\bibitem[Webb(2014)]{Webb14}
Webb, S. D. 2014, J. Comput. Phys., 270, 570

\bibitem[Whipple et al.(1986)]{Whipple86}
Whipple, E. C., Northrop, T. G. \& Birmingham, T. J. 1986, J. Geophys. Res., 91, 4149

\bibitem[Wu et al.(2013)]{Wu13}
Wu, P., Perri, S., Osman, K., et al. 2013, ApJL 763, L30

\bibitem[Zimbardo et al.(2010)]{Zimbardo10}
Zimbardo, G., Greco, A., Sorriso-Valvo, L., et al. 2010, Space Sci. Rev. 156, 89


\end{thebibliography}

\end{document}